\documentclass[journal]{IEEEtran}
\usepackage{amssymb}
\usepackage{amsmath}
\usepackage{booktabs}
\usepackage[ruled,vlined,linesnumbered]{algorithm2e}
\usepackage{amsmath}
\usepackage{amssymb}
\usepackage{bm} % for bold math symbols
\usepackage{textgreek} % for Unicode Greek letters
\usepackage{graphicx}
\usepackage{subcaption}
\usepackage{url}
\usepackage[hidelinks]{hyperref} 
\usepackage{soul}
\usepackage{xcolor}
\usepackage[numbers]{natbib}
\sethlcolor{yellow}

\ifCLASSINFOpdf
\else
\fi
\begin{document}
\title{DoHFuse: A Dual-Branch Architecture with DMAGLSTM for Website Fingerprinting over DNS over HTTPS/3}

\author{
ZunDong Zhang,
Yanan Cheng,
Zhaoxin Zhang,
Xueyang Huo,
Changjiang Wu
\thanks{
Z. Zhang, Y. Cheng, and Z. Zhang are with the Faculty of Computing, Harbin Institute of Technology, China 
(e-mail: 22b903090@stu.hit.edu.cn; chengyn@hit.edu.cn; zhangzhaoxin@hit.edu.cn).
}
\thanks{
X. Huo and C. Wu are with the China Mobile Research Institute, China 
(e-mail: huoxueyang@cmict.chinamobile.com; wuchangjiang@cmict.chinamobile.com).
}
\thanks{
Corresponding author: Yanan Cheng, Zhaoxin Zhang.
}
}


\markboth{Journal of \LaTeX\ Class Files,~Vol.~14, No.~8, August~2015}%
{Shell \MakeLowercase{\textit{et al.}}: Bare Demo of IEEEtran.cls for IEEE Journals}



\maketitle

\begin{abstract}
As personal data privacy becomes increasingly critical in Internet of Things (IoT) environments, secure DNS protocols such as DNS over HTTPS (DoH) and DNS over TLS (DoT) have been widely adopted to protect device communications. However, without effective obfuscation, these protocols remain vulnerable to Website Fingerprinting (WF) attacks that can reveal user activity. With the ongoing deployment of DNS over HTTP/3 (DoH/3) in modern networked systems, padding strategies have been increasingly applied in practice. It is therefore essential to investigate whether DoH/3 can effectively mitigate WF attacks in realistic IoT and edge-network scenarios. To address this, we first collect and publicly release the first real-world benchmark dataset of DoH/3 traffic, generated from domain resolution processes in practical network environments. We further propose DoHFuse, a dual-branch learning framework that integrates inter-arrival time sequences and refined statistical features through an improved DMAG-LSTM, specifically designed to capture burst-aligned temporal patterns. Experimental results show that DoHFuse achieves an accuracy of 88.05\% (precision 88.56, recall 87.96, F1 87.83) in a closed-world setting of 449 classes, and an AUPRC of 0.975 with an F1 score of 0.951 (precision 0.906, recall 1.0) in open-world detection. These findings demonstrate that DoH/3 traffic remains susceptible to WF attacks, suggesting that commonly deployed padding mechanisms alone are insufficient to ensure privacy protection in IoT-scale encrypted DNS communications.
\end{abstract}

\begin{IEEEkeywords}
Dns over Https; Encrypted Traffic; Website fingerprint; privacy risk
\end{IEEEkeywords}

% \fi
\IEEEpeerreviewmaketitle

\section{Introduction}
Website fingerprinting (WF) attacks pose a severe threat to online privacy: by analyzing patterns in encrypted traffic, a passive adversary can infer the sites a user visits, assemble behavioral profiles, and mount targeted phishing or social-engineering campaigns. This threat is particularly concerning in Internet of Things (IoT) environments, where large numbers of devices rely on encrypted communications while often lacking strong endpoint protections. This capability undermines the privacy guarantees users expect from encryption and has kept WF an active research topic for more than a decade.

Most prior WF work has focused on anonymity systems such as Tor and encrypted HTTPS traffic between clients and web servers. With the widespread deployment of encrypted DNS---most notably DNS over HTTPS (DoH) and DNS over TLS (DoT)---DNS resolution itself has emerged as a distinct and increasingly important attack surface \citep{li2023longitudinal,lee2024dohdowngrade}. Against this backdrop, it is natural to ask how WF translates to encrypted DNS, particularly in large-scale IoT and edge-network deployments where DNS traffic is both frequent and latency-sensitive.

Compared to HTTPS, encrypted DNS offers a uniquely advantageous vector for fingerprinting. Queries for a recursive resolver are less contaminated by the complexities of CDN resources, advertisements, and heterogeneous web objects that confound HTTPS-based WF \citep{englehardt2016online}. Feature extraction can thus be cleaner and more directly attributable to a target website. Furthermore, DNS resolution logically precedes the establishment of HTTPS connections, so a successful DNS-based WF attack can provide an earlier signal of user intent, potentially enabling earlier interception or policy enforcement. In IoT scenarios, such early-stage inference can expose device behavior patterns and application usage at scale.

Although numerous studies have shown that DoH, DoT, and DNS over QUIC (DoQ) are vulnerable to fingerprinting in the absence of padding strategies, there is still no systematic investigation of such attacks under the DoH/3 protocol \citep{siby2020encrypted,salari2025privacy}. DoH/3---a modern variant of DoH built atop QUIC---is gaining momentum due to low latency, multiplexing, and connection reuse, with growing deployment in platforms such as Chrome and Cloudflare DNS. Compared with DoH over HTTP/2, DoH/3 introduces new obfuscation of the transport layer and traffic patterns \citep{li2024worldwide}: HTTP/3 (QUIC) supports stream and connection multiplexing, yielding intertwined flows of queries, responses, and control messages (e.g., \texttt{SETTINGS}, \texttt{ACK}, \texttt{PRIORITY\_UPDATE}) \citep{smith2021website}. These characteristics complicate traffic separation and make traditional length-based features far less effective. This creates a critical gap: the feasibility and effectiveness of WF under realistic DoH/3 conditions remain unclear, especially in IoT-scale encrypted DNS deployments. 

\textbf{Research Objectives.} This study aims to:
\begin{enumerate}
  \item Systematically assess DoH/3's susceptibility to WF under realistic HTTP/3/QUIC conditions.
  \item Establish and publicly release a reusable and standardized DoH/3 WF benchmark (data, code, and evaluation templates).
  \item Develop and validate \textbf{DoHFuse}---a dual-branch model with \textbf{DMAG-LSTM}---for robust closed world and open world evaluation.
  \item Provide empirical insights that inform modeling choices, privacy assessment, and defense design, with implications for large-scale IoT traffic analysis.
\end{enumerate}

\textbf{Contributions.} To address these challenges, we conduct a systematic study and make the following key contributions:
\begin{itemize}
  \item We construct and publicly release the first dedicated DoH/3 WF dataset, comprising traffic traces from 449 websites collected over multiple rounds to enable reproducible research.
  \item We propose DMAG-LSTM, a gated LSTM with multi-scale forget-gate fusion tailored to capture the interleaved burst structures prevalent in DoH/3, enhancing temporal modeling.
  \item We design DoHFuse, a dual-branch architecture that softly fuses discriminative statistical features with DMAG-LSTM sequences, improving robustness and generalization.
  \item We characterize DoH/3 traffic by isolating valid request–response pairs and exposing a burst-gap request pattern, laying the groundwork for targeted, timing-aware defenses in encrypted DNS environments.
\end{itemize}

Through extensive experiments in both closed world and open world scenarios with varying class granularities (up to 449 websites), we demonstrate the superior performance and robustness of our approach compared to state-of-the-art baselines under a unified protocol.

\section{Related Work}

\subsection{Website Fingerprinting Attacks on Tor and HTTPS}

Recent studies have advanced website fingerprinting attacks in both Tor and HTTPS traffic. Sirinam et al pioneered the use of deep learning for fingerprinting in their Deep Fingerprinting model, significantly outperforming previous ML-based methods and demonstrating strong robustness even against various defenses \citep{sirinam2018deep}. Cherubin et al. proposed a dynamic expansion architecture for class-incremental WF in real-world Tor settings \citep{cherubin2022online}. Shen et al. introduced Robust Fingerprinting using a traffic aggregate matrix to build resilient traffic representations against defenses \citep{shen2023subverting}. Abolfathi et al. developed SuperLearner, an ensemble-based model that achieved up to 97. 2\% accuracy in HTTPS traffic \citep{abolfathi2024toward}. Jansen, Wails, and Johnson review the real-world evaluation of WF by reconstructing labeled entry-side traces from genuine Tor traffic ~\citep{jansen2024retracer}. Despite these advances, existing models remain highly data-dependent and struggle to generalize to dynamic or updated websites, limiting their real-world applicability.

Methodological advances in WF include signal-efficient attacks and low-sample generalization. 
Attarian and Keshavarz-Haddad show that even the first-packet direction can be remarkably informative \citep{attarian2023firstpacket}; the few shot WF is explored by Chen et al. \citep{chen2021fewshot} and extended the cross-domain via Brownian distance covariance by Zou et al.\ \citep{zou2022crossdomain}. 
Zhan et al.\ study WF on \emph{early} QUIC traffic \citep{zhan2021earlyquic}. 
However, most of these techniques were developed for Tor/HTTPS or for DNS settings without robust padding and HTTP/3 multiplexing. 

\subsection{Encrypted Malicious-Traffic and TLS Fingerprinting}
Beyond DNS, several works analyze encrypted malicious traffic at the transport layer. 
Yu et al. propose a TLS fingerprinting approach with an attributed graph kernel \citep{yu2024tls}, 
and Hong et al. use graph-based multiview features for encrypted malicious traffic detection \citep{hong2023graph}. 
In parallel, Keshk et al. introduce an explainable deep learning intrusion detection framework for IoT networks \citep{keshk2023xai}, 
and earlier traffic classification systems explore dynamic multiclass pipelines \citep{xiao2019dynamic}. 
These directions inform feature modeling and interpretability, but differ from our objective: they target malware/attack detection on TLS/IoT traffic rather than DoH/3 website fingerprinting.

\subsection{Encrypted DNS Detection and DNS-based Security}
A complementary line of work focuses on \emph{detecting} encrypted DNS usage or DNS-based malicious activity, rather than inferring website identity. Crucially, these methods help \textbf{separate DoH flows from background HTTPS traffic}, providing the pre-filtering and labeling capability our study also relies on. Jerabek et al. comparatively evaluate DoH detectors \citep{jerabek2024comparative}; Zhan et al.\ detect data exfiltration over DoH \citep{zhan2022exfiltration}; and Quezada et al.\ build a real-time bot infection detector using DNS fingerprinting and ML \citep{quezada2023realtime}. Although these systems primarily target DoH/DoT (not DoH/3’s HTTP/3-over-QUIC setting), they offer practical \emph{front-end} tools, DoH-vs-HTTPS separation and anomaly flags, that act as prerequisites to our DoH/3 WF pipeline (e.g., curated trace selection and ground truth preparation). In contrast, our work asks: \emph{once DoH/3 flows are isolated}, can site identity still be inferred under padding and multiplexing?

\subsection{Website Fingerprinting Attacks on Encrypted DNS}

The growing deployment of encrypted DNS protocols such as DoH and DoQ has prompted extensive research into their vulnerability to website fingerprinting (WF). For example, Bushart and Rossow systematically evaluated padding-based countermeasures in encrypted DNS and found that simple padding is insufficient to thwart WF attacks \citep{bushart2020padding}. Similarly, Hynek and Cejka showed that unpadded DoH leaks domain information through packet length correlations, creating a 'privacy illusion' unless padding mechanisms are applied\citep{hynek2020privacy}. Siby et al. demonstrated that even encrypted DNS queries can leak significant identifying information through observable traffic patterns, particularly when packet length features are not adequately obfuscated \citep{siby2020encrypted}. Mazzuz and Shabtai showed that even with domain name encryption, website fingerprinting remains effective, proposing a Siamese network–based few-shot attack that achieves high accuracy with limited samples \citep{mazzuz2025domain}.

Many studies have shown that packet length-related characteristics are among the most critical factors in website fingerprinting. Zou et al. proposed DePL, which uses encrypted traffic packet length feature sequences and employs a BiLSTM model to detect privacy leakage in DNS-over-HTTPS \citep{zou2021depl}. Trevisan et al. demonstrated that, despite using DoH and ECH, traffic analysis based on encrypted flow features can still effectively infer user-visited services, revealing residual privacy leakage risks \citep{trevisan2023attacking}. Hu and Fukuda analyzed DNS-over-QUIC traffic and showed that packet length and timing features can leak queried domain information, proposing padding-based countermeasures to mitigate such privacy risks \citep{hu2024privacy}.

Several works have explored timing-based WF under encrypted DNS protocols. Shao et al. proposed a lightweight feature extraction approach based on interpacket intervals and showed competitive results using traditional classifiers like Random Forests \citep{shao2023lightweight}. In contrast, Dahanayaka et al. investigated inline traffic analysis attacks on DoH, highlighting the vulnerability of passive observers to extract features from QUIC traffic streams \citep{dahanayaka2022inline}.

Recent advances in machine learning have also been applied to encrypted DNS traffic analysis. Salari et al. proposed a hybrid deep learning approach that combines statistical features and sequence models, achieving improved accuracy over previous baselines when evaluating oblivious DoH (ODoH) \citep{salari2025privacy}. Meanwhile, Csikor et al. introduced Transformer-based models for privacy analysis on HTTP/3 and DoQ traffic, highlighting the need for adaptive architectures that can capture fine-grained temporal dynamics \citep{csikor2025dns}.

Although these efforts provide valuable information, most of them rely heavily on the length or size of the packet as a primary discriminative feature. As shown in our experiments and consistent with the findings in \citep{siby2020encrypted}, such characteristics become significantly less effective in the presence of EDNS(0) padding and QUIC-level obfuscation, which are prevalent in DoH/3.

Furthermore, although previous work has assessed website fingerprinting in DoH, DoT, and DoQ, a systematic evaluation targeting DoH/3 is still lacking. In summary, while previous literature has explored website fingerprinting across various encrypted DNS protocols, our work is among the first to offer a dedicated dataset, a customized temporal modeling architecture, and a systematic evaluation for DoH/3 fingerprinting.

\section{Preliminaries‌}
\subsection{Motivation and Current Status}
DNS over HTTPS encapsulates DNS messages inside HTTPS requests and responses, providing end-to-end confidentiality and integrity \citep{hoffman2018rfc8484}. This design prevents on-path eavesdropping and tampering and limits inspection or manipulation that relies on plaintext DNS.

With the maturation of HTTP/3 (RFC 9114) over QUIC (RFC 9000), DoH can run atop HTTP/3—referred to in this paper as DoH/3 \citep{bishop2022rfc9114,iyengar2021rfc9000}. DoH/3 benefits directly from QUIC’s properties: a 1-RTT TLS 1.3 handshake (and 0-RTT for session resumption), built-in stream multiplexing, and resilience to head-of-line blocking, which together reduce latency and improve robustness.

On the resolver side, major public resolvers already support DoH and expose secure transports suitable for HTTP/3. Google Public DNS documents DoH with modern TLS 1.3 practices and QUIC/HTTP/3 paths, and recommends EDNS(0) padding (RFC 7830/8467) to shrink the attack surface of size-based traffic analysis \citep{mayrhofer2016rfc7830,mayrhofer2018rfc8467}. Cloudflare’s 1.1.1.1 similarly supports DoH over HTTP/2 and HTTP/3 with multiplexing, encouraging at least HTTP/2 and taking advantage of HTTP/3 when available\citep{cloudflare_doh_api_2025}.

On the client side, Chrome has shipped HTTP/3/QUIC support broadly\citep{chromium_quicer_2024}, so when a resolver advertises HTTP/3 the browser can issue DoH queries over HTTP/3. Android has also integrated encrypted DNS with support for DoH/3 (initially targeting well-known providers such as Google and Cloudflare)\citep{maurer_doh3_android_2022}. 
\subsection{DoH/3 Connection Establishment and Query Flow}

\begin{figure}
    \centering
    \includegraphics[width=1\linewidth]{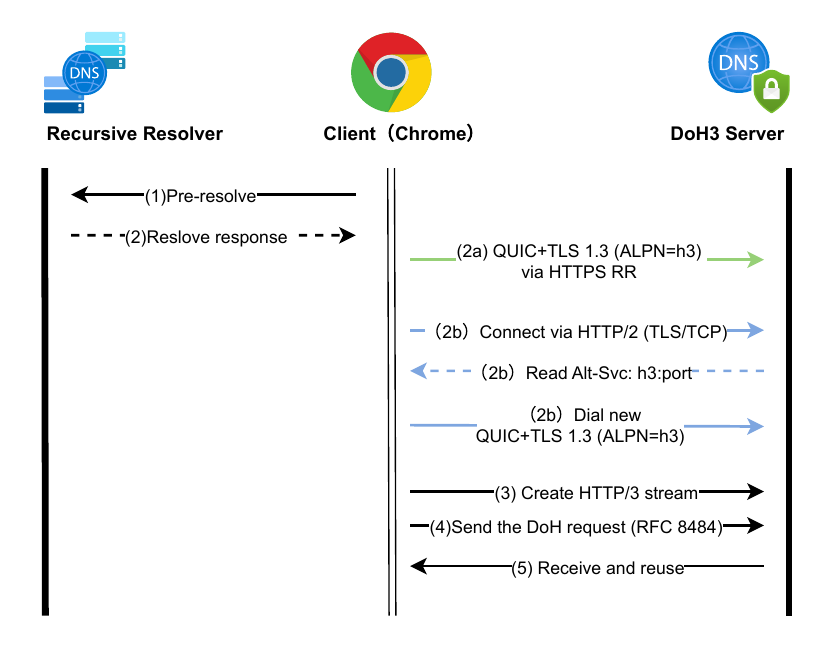}
    \caption{DoH/3 connection establishment schematic.}
    \label{fig:doh3-establish}
\end{figure}

Figure \ref{fig:doh3-establish} summarizes the establishment of DoH/3 and the
query flow between the client (e.g. Chrome) and a public resolver. 

\textbf{(1)Pre-resolve.} Send a conventional DNS query to the recursive resolver configured in the system (typically provisioned by DHCP), requesting the \textbf{A}, \textbf{AAAA}, and \textbf{HTTPS (SVCB/HTTPS, type 65)} \citep{schwartz2023rfc9460} records for the DoH hostname (e.g., \texttt{dns.google.com}).

\textbf{(2) Resolver response.} Receive DNS responses from the system-configured recursive
resolver; responses include \textbf{A/AAAA} and may include an \textbf{HTTPS (SVCB/HTTPS, type~65)}
record for the DoH hostname.

\textbf{(2a) HTTPS RR present.} If an HTTPS record is present, consume the advertised parameters (e.g., \texttt{alpn="h3,h2"}, \\ \texttt{ipv4hint/ipv6hint}, ECHConfig) and establish \\
\textbf{QUIC + TLS~1.3} (\textbf{ALPN}=\texttt{h3}) directly to the indicated endpoint.\,
(See the HTTPS-RR note for a canonical example.)

\noindent
\fbox{%
  \begin{minipage}{0.97\linewidth}\small
  \textbf{HTTPS RR example}\\[-2pt]
  \texttt{cloudflare-dns.com.\quad 65\quad IN\quad HTTPS\quad 1\quad .\quad
  alpn="h3,h2"\quad
  ipv4hint=104.16.248.249,104.16.249.249\quad
  ipv6hint=2606:4700::6810:... ,\;2606:4700::6810:...}
  \end{minipage}
}

\textbf{(2b) No HTTPS RR.} If no HTTPS record is present, first connect over \textbf{HTTP/2}
(\textbf{TLS}/TCP) to the DoH endpoint and read the server’s \textbf{Alt-Svc: h3}; if accepted by
policy, initiate a new \textbf{QUIC + TLS~1.3} connection (\textbf{ALPN}=\texttt{h3}) and proceed
over HTTP/3. \emph{Note:} Alt-Svc announces an alternative endpoint rather than upgrading the
existing TCP connection.

\textbf{(3) Establish HTTP/3 session.} After the handshake (typically 1-RTT; 0-RTT on resumption), create an HTTP/3 stream (headers encoded with QPACK).

\textbf{(4) Send the DoH request (RFC 8484).} Use either 
\emph{GET} \path{/dns-query?dns=BASE64URL(dns-wire, no padding)} with \path{Accept: application/dns-message}, or \emph{POST} \path{/dns-query} carrying the raw DNS wire message with 

\path{Content-Type: application/dns-message}; \emph{POST} 
offers better privacy, while \emph{GET} remains cache-friendly.

\textbf{(5) Receive and reuse.} Receive 200 OK (application/dns-message) and multiplex subsequent queries on the same HTTP/3 connection.

In our measurements, Google Public DNS does not publish an HTTPS record for its DoH endpoint, whereas Cloudflare DNS does publish an HTTPS record. In both cases, Chrome issues DoH queries using POST to /dns-query with \texttt{Content Type: application/dns-message}, and the HTTP body carries the raw DNS wire message. We observe that this payload includes EDNS(0) padding (per RFC 7830/8467), which helps obscure message size and mitigates length-based traffic fingerprinting.

\section{Method}
\subsection{Threat Model}
\begin{figure}
    \centering
    \includegraphics[width=1\linewidth]{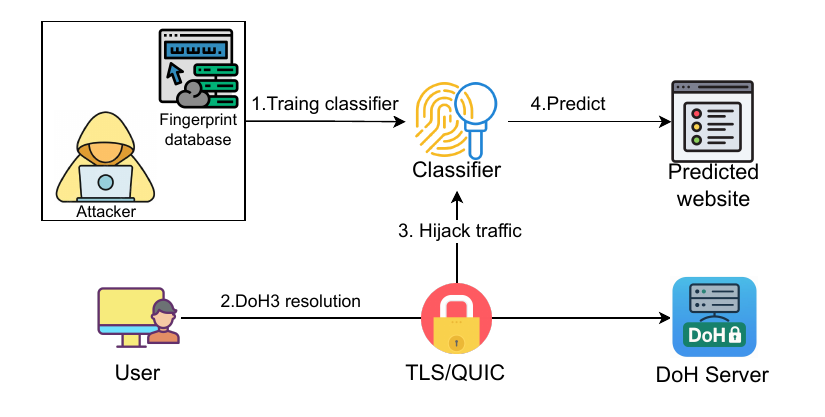}
    \caption{Threat model for DoH/3 website fingerprinting.}
    \label{fig:threatmodel}
\end{figure}

Our threat model, depicted in Figure \ref{fig:threatmodel}, follows the traditional approach in WFP attacks. The adversary is a passive on-path observer between the client and the DoH/3 resolver: able to monitor traffic but unable to inject, drop, or modify packets and unable to decrypt TLS/QUIC. Observable features include the 5-tuple, connection start/finish times, and the QUIC packet sequence (timestamps, directions, sizes). Flows are coarsely segmented by 5-tuple and idle timeouts.

We study whether an adversary can identify user visits solely from DoH/3 traffic. The adversary first gathers in\mbox{-}situ measurements in the target network environment (e.g., the same geographic or ISP region as the victim). Specifically, the adversary visits a set of websites of interest over DoH/3 and records the resulting packet traces, which serve as labeled data to train a supervised classifier.

After training, the adversary applies the model to real-world traffic traces collected from passive observation. Given a new DoH/3 flow, the classifier analyzes its observable features and attempts to predict the most likely target website. If successful, this reveals the user’s browsing behavior without decrypting any content, demonstrating a privacy vulnerability even in the presence of strong encryption of the transport layer.

\subsection{Data Collection}

To enhance the robustness and generalizability of our analysis, we collected DoH/3 traffic from two geographically distinct vantage points in the United States during \emph{[July 2025]}. Specifically, we deployed Virtual Private Servers (VPS) located in \textbf{Los Angeles, California} and \textbf{San Jose, California}, each configured as an independent client host. Both instances ran Ubuntu~24.04 on x64 hardware with identical default network configurations (the same kernel, TCP stack parameters and DNS system settings). The same Selenium-based crawling script and identical versions of Chromedriver (v141.0) and Google Chrome (v.141.0) were executed across both vantage points to ensure consistency in client behavior.

% We collected DoH/3 traffic from two geographically distinct VPS vantage points during \emph{[2025--07]}, in order to enhance the generalization of the models developed. These clusters consisted of x86 servers running Ubuntu 24.04 stock
% with default network configurations.

Candidate domains were drawn from two established rankings: Majestic Million (top~400) and Tranco (top~200)~\citep{MajesticMillion,pochat2018tranco}. After merging and deduplication, we obtained \textbf{449} unique domains. Entries corresponding to CDN/API endpoints or persistently unreachable names (e.g. blocked by firewalls) were excluded from the target set.

Each reachable domain was visited 100 times using a Selenium automation framework with Chromedriver that controls headless Chrome. This resulted in approximately 44,900 valid sessions after filtering out failures. A new Chrome profile and process were instantiated for each visit to prevent DNS caching. All DNS queries were forced to use DoH/3 via \texttt{dns.google.} Traffic was recorded with \texttt{tcpdump}, and TLS session keys were logged via SSLKEYLOGFILE for optional validation, although our features do not require decryption.

For closed world evaluation, we use all 449 domains. For open world tests, we additionally sampled domains ranked 500--600 in Majestic (ten visits per domain) to form a background set unseen during training.

Regarding trace scope, for 226 domains we retained \emph{full-session} captures (entire page load); for the remaining domains we retained \emph{DoH-only} captures by filtering resolver-bound UDP/443 traffic to known DoH addresses.

We release both the \emph{raw packet captures (PCAP)} and the \emph{derived per-visit CSV statistics} produced by our feature pipeline (e.g. request counts and timing summaries). These artifacts are provided in the accompanying data set repository so that other researchers can directly reuse the raw traces or the precomputed statistics without replicating the entire parsing stack. All traffic comes from our own clients visiting public websites; no third-party user data was collected. The dataset is available on Scidb.\footnote{\url{https://www.scidb.cn/anonymous/clFCUkZ2}\; (anonymous link for double-blind review).}

\subsection{DoHFuse Overall Architecture}
\begin{figure}[h!t]
    \centering
    \includegraphics[width=1\linewidth]{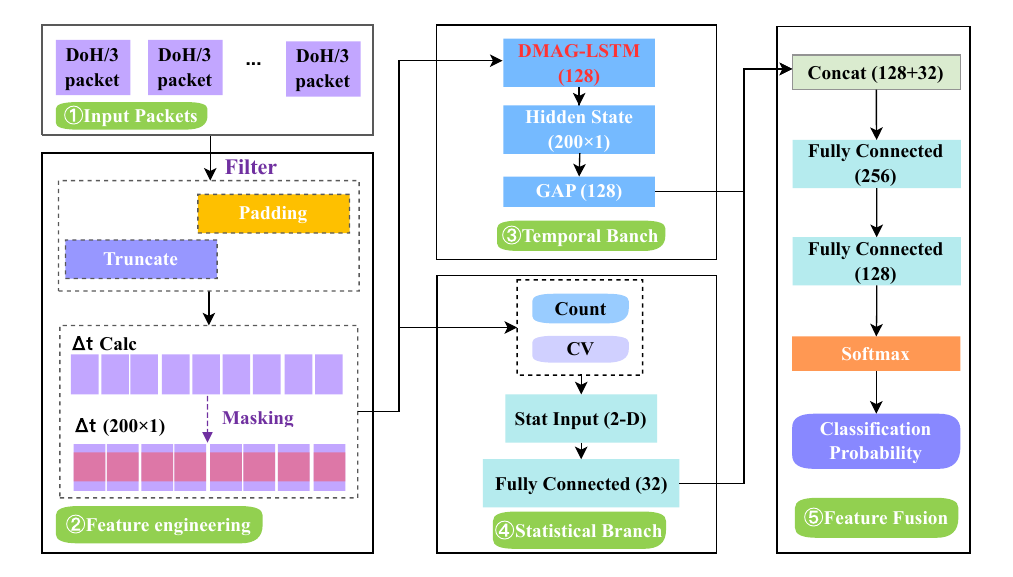}
    \caption{Overview of the proposed DoHFuse architecture. The model integrates a timing-sensitive temporal branch based on DMAG-LSTM and a statistics-based branch for coarse traffic summarization.}
    \label{fig:arch}
\end{figure}

To provide a clearer understanding of our system design, the following section presents the overall architecture of DoHFuse, outlining the relationship between its feature inputs, model training, and classification pipeline. Figure~\mbox{\ref{fig:arch}} presents the overall architecture of the proposed DoHFuse model. The framework is designed to jointly leverage \emph{fine-grained temporal request patterns} and \emph{coarse-grained statistical traffic features} to enhance robustness under diverse DoH/3 conditions. In HTTP/3-over-QUIC, encryption, padding, and multiplexing obscure size cues while producing interleaved \emph{burst--gap} timing. Hence, DoHFuse adopts a \emph{timing-first, dual-branch} design that fuses short-term burst dynamics and long-term statistical regularities, aligning more naturally with transport-layer realities than length-centric pipelines.

As shown in Figure~\ref{fig:arch}, DoHFuse is organized into five sequential components:
(1) \emph{Input (Packets)}, (2) \emph{Feature Engineering}, 
(3) \emph{Temporal Branch}, (4) \emph{Statistical Branch}, and 
(5) \emph{Feature Fusion}. 
This modular structure provides a clean processing flow from raw QUIC packets to the final 
classification output while separating heterogeneous timing and statistical signals.

\textbf{(1) Input (Packets)}  
The first stage of our pipeline converts raw DoH/3 traffic into a structured CSV format suitable
for later processing. Each website visit was originally recorded as a QUIC pcap trace. From this
trace, we isolate only DoH/3-related packets by filtering UDP traffic on port~443 and
retaining flows whose source or destination corresponds to known DoH resolvers. For every
retained packet, we extract its arrival timestamp, compute the relative time (offset from the
first DoH packet of the visit), calculate the inter-packet interval, and record essential
metadata such as IP addresses, ports, packet length, and direction.

These packet-level attributes are written sequentially into a per-visit CSV file, where each row
represents one DoH/3 packet and each column corresponds to a normalized field required for the
subsequent feature engineering stage. The CSV files produced in this step form the actual input to DoHFuse.

Although this conversion is performed offline in our dataset pipeline, the same filtering and
extraction logic can be applied to a live capture stream, enabling real-time generation of CSV
traces for online analysis.

\textbf{(2) Feature Engineering.}  
From the filtered DoH/3 packet stream, we compute two complementary feature groups:
a sequence of inter-request intervals (Δt) and a set of coarse statistical attributes.
Operations such as truncation, padding, and masking are applied to standardize the sequence length (200×1). Since this step involves several domain-specific operations tailored to DoH/3 timing, the complete formulation and justification are presented separately in Section~\ref{sec:feature}.

\textbf{(3) Temporal Branch.}  
The Δt sequence is processed by a dedicated temporal encoder. After masking variable-length inputs, the sequence is fed into the proposed \textbf{DMAG-LSTM} module (128 units), followed by normalization, regularization, and global average pooling to obtain a compact temporal embedding. DMAG-LSTM is designed to capture both micro-burst and long-gap timing structures that naturally arise in multiplexed DoH/3 traffic. Because the internal computation and the relationship between DMAG-LSTM and standard LSTM are central contributions of this work, they are detailed later in Section~\ref{sec:dmag}.

\textbf{(4) Statistical Branch.}  
In parallel to the temporal encoder, the statistical branch operates on two global indicators
derived from each DoH/3 trace: the total number of client-to-resolver request packets (\emph{query count}) and the coefficient of variation (\emph{CV}) of their inter-packet
intervals. These variables capture complementary aspects of the DoH/3 behavior. The query count
reflects the overall intensity of DNS activity triggered by a webpage, which correlates with its
resource structure and is largely unaffected by padding or frame coalescing. 

Together, these two statistics provide a coarse but reliable description of the global request shape, serving as padding-invariant, low-variance signals that compensate for the fine-grained sensitivity of the temporal branch. They are fed into a lightweight feedforward block consisting of a dense projection layer with ReLU activation, BatchNormalization, and Dropout regularization, expanding the two inputs into a 32-dimensional embedding. This embedding acts as a global prior that stabilizes training, constrains the fused representation, and improves generalization when the number of classes scales up or when timing variability increases across vantage points.

\textbf{(5) Feature Fusion and Classification.}  
The outputs of the two branches are concatenated to form a unified latent representation.
This fused vector is then processed through a lightweight classifier composed of:
\begin{itemize}
    \item Dense (256 units, ReLU) + BatchNormalization + Dropout
    \item Dense (128 units, ReLU)
    \item Output Dense layer producing class logits
\end{itemize}
Training uses categorical cross-entropy loss with balanced class weighting to mitigate label imbalance.
A \emph{late soft fusion} strategy is adopted to allow each branch to preserve its inductive bias (local timing vs. \ global traffic shape) before integration.
This design avoids modality dominance and yields a calibrated fused space suitable for both large-class closed-world recognition and open-world rejection.

Following the fusion stage, DoHFuse employs a structured training procedure to ensure stable optimization of both branches. In addition to the architectural components described above, DoHFuse adopts a structured training pipeline that is conceptually aligned with its dual-branch design. The optimization process follows a two-stage procedure. In the first stage, the model is trained with an adaptive learning-rate schedule and early stopping to stabilize both branches and ensure that the temporal and statistical representations converge to consistent scales. In the second stage, a low learning-rate fine-tuning step is applied to refine the fused representation and sharpen the decision boundary. 

Overall, these five components together form a coherent dual-branch framework that remains robust under the timing distortions, padding behavior, and multiplexing characteristics inherent to DoH/3 traffic. Detailed feature engineering procedures are discussed in Section~\ref{sec:feature}, while the DMAG-LSTM module is analyzed 
in depth in Section~\ref{sec:dmag}.

\subsection{Feature engineering}
\label{sec:feature}

By observation, we found that DoH/3 traffic contains a significant volume of QUIC-layer control frames that are not related to DNS requests or responses. Consequently, these must be excluded during feature extraction. Due to EDNS(0) padding and HTTP/3 framing (QPACK), DoH/3 request and response payloads exhibit highly concentrated size distributions. Specifically, more than 99\% of the request packets fall within the range of 233–350 bytes, while more than 99\% of response packets lie between 555–777 bytes. Within these ranges, non-query and non-response packets are extremely rare. This distribution can be attributed to the use of padding and the multi-stream transmission pattern. Each query is carried over at least two streams: the first is STREAM X (where X is even), which contains the DNS message and is padded to a fixed size due to EDNS(0); the second is STREAM 2, which sets the priority of STREAM X to 1. In a few cases, additional streams may appear to convey auxiliary information, but their contribution is negligible. A similar mechanism applies to response packets, which also involve at least two streams: STREAM Y (carrying the DNS response) and STREAM 11, which are used for header compression references that facilitate client-side interpretation. This streaming strategy explains why queries and responses are confined to narrowly defined size ranges.

This clear separation facilitates the straightforward extraction of request and response datasets. However, it simultaneously limits the versatility of packet length features for broader classification tasks, rendering them ineffective for model training.

Although requests and responses typically form pairs of one-to-one, network conditions introduce greater variability in response timing and sequencing. The client-initiated request timing exhibits comparatively higher stability. Therefore, we exclusively extract features from uplink packets (client → resolver).

For encrypted traffic, the selection of features is inherently constrained, generally falling into three categories: timing, counts, and packet size\citep{niakanlahiji2023toward}. Given size obfuscation through padding, timing-based features become the primary dependency. Accordingly, we selected inter-request intervals as the key discriminative feature:

\[
\Delta t_k = t_k - t_{k-1},\quad k=2,\dots,n,\quad \Delta t_1=0.
\]

Here, $t_k$ denotes the timestamp at which the $k$-th request packet is sent from the client, measured relative to the start of the capture session. The inter-request interval $\Delta t_k$ therefore represents the elapsed time between two consecutive request packets, with $\Delta t_1$ defined as zero for initialization. To ensure uniform input dimensionality for the model, each interval sequence is truncated or padded to a fixed length $L$ (e.g., $L=200$). Missing positions are assigned a sentinel value of $-1$ and subsequently masked during training (\texttt{mask\_value=-1}) so that they do not contribute to gradient updates.

In addition, two statistical features are incorporated as auxiliary inputs: the total packet count and burstiness, quantified by the coefficient of variation (CV).

\[
\text{count}=n,\qquad
\text{CV}=\frac{\mathrm{std}(\{\Delta t_k\}_{k\ge2})}{\mathrm{mean}(\{\Delta t_k\}_{k\ge2})+\varepsilon},
\]

where $\varepsilon$ avoids division by near-zero means. These two features are selected because of their ability to reflect overall flow activity and temporal irregularity. In particular, the coefficient of variation (CV) characterizes the burst of query intervals ---- a key discriminative trait in DoH/3 traffic ---- by capturing the extent of timing fluctuation between adjacent packets \citep{goh2008burstiness}.

The resulting input consists of a length-$L$ sequence
$\mathbf{x}^{(\text{seq})}\in\mathbb{R}^{L}$ and a 2-D vector
$\mathbf{x}^{(\text{stat})}\in\mathbb{R}^{2}$ that feed the temporal
and statistical branches of our model, respectively.

\subsection{DMAG-LSTM}
\label{sec:dmag}

% The temporal branch of DoHFuse is designed to model the characteristic burst–gap structure of DoH/3 request sequences. Standard LSTM architectures, although effective for generic sequential modeling, rely on a \emph{single} forget gate and a \emph{single} input modulation path. This makes them prone to either oversmoothing dense micro-bursts or forgetting longer-range timing dependencies across gaps. To address these limitations, we introduce the \textbf{DMAG-LSTM} (Dual-scale Modulated and Adaptive-Gated LSTM), which preserves the core LSTM update structure while enhancing its gating mechanisms in three ways:
% (1) multi-scale forget-gate fusion, (2) dynamic input modulation, and (3) CIFG-based gate simplification. Fig.~\ref{fig:DMAGLSTM} illustrates how these modifications are incorporated into the standard LSTM cell.

Long Short-Term Memory (LSTM) is a recurrent neural network architecture designed to model sequential data while mitigating the vanishing-gradient problem in long sequences. At each time step, a standard LSTM maintains a hidden state and a memory cell, and updates them through three gates: the forget gate, which controls how much past memory is retained; the input gate, which determines how much new information is written into the cell; and the output gate, which regulates how much cell information is exposed to the hidden state. This gating mechanism allows LSTM to capture both short-term variations and longer-range dependencies in temporal sequences.

In our setting, LSTM provides a natural baseline for modeling inter-request interval sequences in DoH/3 traffic. However, DoH/3 traces often exhibit a mixture of tightly clustered micro-bursts and relatively stable inter-burst gaps. Under such burst--gap patterns, a standard single-scale gating scheme may either smooth out fine-grained burst structures or fail to retain useful information across longer gaps. This motivates the design of our \textbf{DMAG-LSTM} (\textbf{D}ual-scale \textbf{M}odulated and \textbf{A}daptive-\textbf{G}ated LSTM).

\begin{figure}[!ht]
    \centering
    \includegraphics[width=1\linewidth]{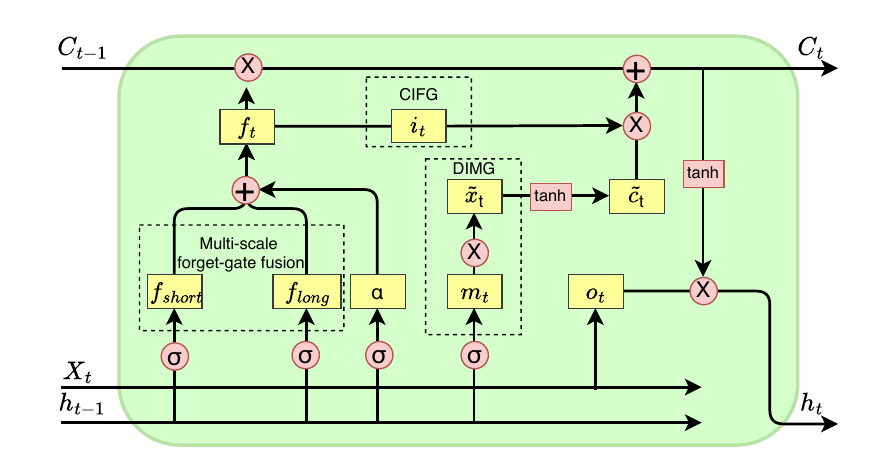}
    \caption{DMAGLSTM: Internal structure of the proposed multi-scale gated LSTM cell.}
    \label{fig:DMAGLSTM}
\end{figure}

DMAG-LSTM preserves the standard LSTM recurrence, including hidden-state propagation and cell-state update, but refines several internal gating operations to better match the temporal dynamics of encrypted DoH/3 traffic. As illustrated in Fig.~\ref{fig:DMAGLSTM}, the proposed cell introduces three main modifications:
(1) multi-scale forget-gate fusion,
(2) dynamic input modulation, and
(3) CIFG-based gate simplification.

% Before describing the extensions, we briefly recall the standard LSTM update:
% \[
% \begin{aligned}
% i_t &= \sigma(W_i [h_{t-1}, x_t] + b_i), \\
% f_t &= \sigma(W_f [h_{t-1}, x_t] + b_f), \\
% o_t &= \sigma(W_o [h_{t-1}, x_t] + b_o), \\
% \tilde{C}_t &= \tanh(W_c [h_{t-1}, x_t] + b_c), \\
% C_t &= f_t \odot C_{t-1} + i_t \odot \tilde{C}_t, \\
% h_t &= o_t \odot \tanh(C_t).
% \end{aligned}
% \]

DMAG-LSTM preserves this fundamental structure but replaces—and in some cases merges
, several internal gates to better align with the temporal characteristics of encrypted DoH/3 traffic.

\subsubsection*{Multi-scale forget-gate fusion}

\begin{figure}
    \centering
    \includegraphics[width=1\linewidth]{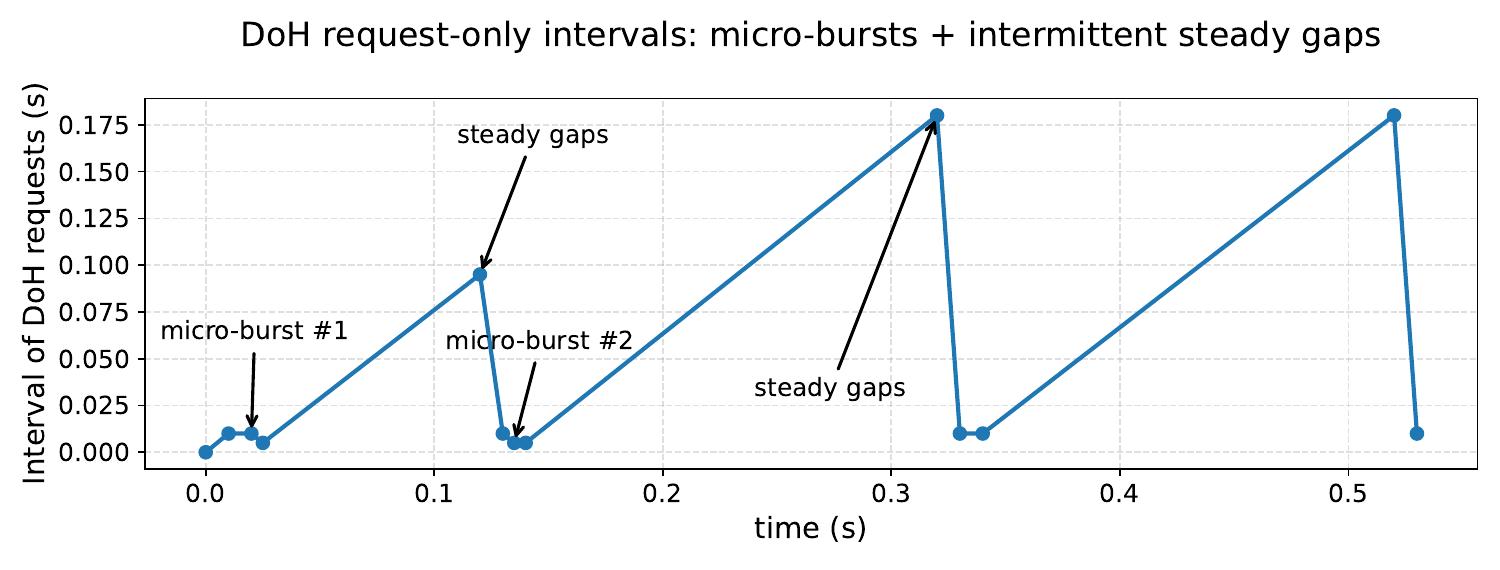}
    \caption{An illustrative example of DoH/3 request-only intervals, showing characteristic micro-bursts and intermittent steady gaps.}
    \label{fig:dohrequest}
\end{figure}

As illustrated in Fig.~\ref{fig:dohrequest}, many DoH/3 request sequences exhibit a characteristic pattern composed of several short micro-bursts, each containing tightly clustered queries issued in parallel to multiple domains during page loading. Between these bursts, we often observe intermittent and relatively stable timing gaps. Such dual-scale behavior---dense intra-burst activity versus sparse inter-burst intervals---poses a challenge for a conventional LSTM with a single forget gate, which must simultaneously suppress local noise and preserve longer-range temporal dependencies.

To address this issue, we introduce two separate forget gates: a \textbf{short-term forget gate} \( f_t^{(s)} \) and a \textbf{long-term forget gate} \( f_t^{(l)} \):
\begin{align}
z_t &= [h_{t-1},\,x_t], \\
f_t^{(s)} &= \sigma(W_f^{(s)} z_t + b_f^{(s)}), \\
f_t^{(l)} &= \sigma(W_f^{(l)} z_t + b_f^{(l)}),
\end{align}

where \( x_t \in \mathbb{R} \) denotes the input inter-request interval at time \( t \), and \( h_{t-1} \in \mathbb{R}^{h} \) is the previous hidden state. The short-term forget gate is initialized with bias \( b_f^{(s)} = -1 \), encouraging faster forgetting of transient fluctuations within local bursts. In contrast, the long-term forget gate is initialized with \( b_f^{(l)} = +1 \), promoting stronger retention of dependencies across burst boundaries.

To adaptively balance these two temporal scales, we introduce a \textbf{fusion gate} \( \alpha_t \in [0,1] \), which is dynamically computed from the current input and hidden state. The resulting forget gate is defined as
\begin{align}
\alpha_t &= \sigma(W_{\alpha} z_t + b_{\alpha}), \\
f_t &= \alpha_t \cdot f_t^{(s)} + (1 - \alpha_t) \cdot f_t^{(l)}.
\end{align}

This adaptive fusion yields a context-sensitive forget gate that can switch between fine-grained temporal sensitivity and coarse-grained memory retention. Importantly, the resulting \( f_t \) remains a drop-in replacement for the conventional LSTM forget gate and therefore preserves compatibility with the standard cell-state update rule.

\subsubsection*{Dynamic Input Modulation Gate (DIMG)}

To enhance the flexibility of the LSTM cell in handling varying traffic rhythms, we introduce a \textbf{dynamic input modulation gate} \( m_t \in [0,1] \), which adaptively controls the influence of current input \( x_t \) on candidate memory content.

Unlike conventional LSTM designs where the input gate solely determines the contribution of \( x_t \), our formulation adds a modulation term that explicitly conditions the relative magnitude of \( x_t \) with respect to the recent history. Formally, the modulated input is computed as follows:

\begin{align}
m_t &= \sigma(W_m z_t + b_m), \\
\tilde{x}_t &= m_t \cdot x_t,
\end{align}

where \( z_t = [h_{t-1}, x_t] \) is the combined input and hidden state. The modulated input \( \tilde{x}_t \) replaces \( x_t \) in the calculation of the candidate memory cell \( \tilde{C}_t \). This allows the network to dynamically suppress or amplify the current input signal based on its temporal context.

In the context of DoH/3 traffic, where input intervals \( x_t \) can alternate between sparse background pulses and dense bursts, this modulation helps the model to better attenuate noise or emphasize salient timing cues. For example, queries to auxiliary Google domains often appear between major bursts and can be down-weighted, while tightly packed DoH requests in a burst receive stronger emphasis.

This mechanism improves the model's discriminative capacity for subtle, yet temporally meaningful, differences in DoH access patterns.

\subsubsection*{Gate Simplification via Coupled Input and Forget Gate (CIFG)}

To reduce parameter redundancy and stabilize training, we adopt the \textbf{Coupled Input and Forget Gate (CIFG)} mechanism \citep{greff2016lstm}. Instead of learning separate input and forget gates, CIFG ties them together through a simple coupling:

\begin{align}
i_t = 1 - f_t
\end{align}

where \( i_t \) is the input gate. This reduces the number of parameters and ensures a complementary balance between memory retention and update.

Given that our input is a 1-D sequence of DoH/3 request intervals, the CIFG strategy introduces no performance bottleneck while improving generalization and simplifying the temporal gating logic.

\subsubsection*{Summary of Relationship to Standard LSTM}

\begin{algorithm}[h!t]
\caption{Forward pass of the proposed DMAG-LSTM cell}
\label{alg:dmaglstm}

\KwIn{Input interval $x_t$, previous hidden state $h_{t-1}$, previous cell state $C_{t-1}$}
\KwOut{Updated hidden state $h_t$, updated cell state $C_t$}

\tcp{Concatenate input and previous hidden state}
$z_t \gets [h_{t-1}, x_t]$ \\

\BlankLine
\tcp{Multi-scale forget-gate fusion}
$f_t^{(s)} \gets \sigma(W_f^{(s)} z_t + b_f^{(s)})$ \\
$f_t^{(l)} \gets \sigma(W_f^{(l)} z_t + b_f^{(l)})$ \\
$\alpha_t \gets \sigma(W_\alpha z_t + b_\alpha)$ \\
$f_t \gets \alpha_t \cdot f_t^{(s)} + (1-\alpha_t)\cdot f_t^{(l)}$ \\

\BlankLine
\tcp{Coupled Input–Forget Gate (CIFG)}
$i_t \gets 1 - f_t$ \\

\BlankLine
\tcp{Dynamic input modulation}
$m_t \gets \sigma(W_m z_t + b_m)$ \\
$\tilde{x}_t \gets m_t \cdot x_t$ \\

\BlankLine
\tcp{Cell state update and output gate}
$\tilde{C}_t \gets \tanh(W_c[h_{t-1},\tilde{x}_t] + b_c)$ \\
$C_t \gets f_t \odot C_{t-1} + i_t \odot \tilde{C}_t$ \\
$o_t \gets \sigma(W_o z_t + b_o)$ \\
$h_t \gets o_t \odot \tanh(C_t)$ \\

\BlankLine
\Return{$h_t, C_t$}

\end{algorithm}

DMAG-LSTM is not an entirely new recurrent architecture independent of LSTM. Rather, it is an \emph{LSTM-compatible extension} that preserves the standard recurrence while refining selected gating operations. Specifically, the forget gate is enhanced through dual-scale fusion, the input pathway is adaptively modulated, and the input--forget interaction is simplified through CIFG. In contrast, the cell-state update rule and the output gate remain unchanged.

Therefore, DMAG-LSTM retains structural compatibility with a standard LSTM while improving its ability to model the multi-scale burst--gap timing patterns that characterize DoH/3 request sequences. The forward procedure is summarized in Algorithm~\ref{alg:dmaglstm}. In implementation, DMAG-LSTM is realized as a custom recurrent layer in PyTorch and directly replaces the standard LSTM block in the temporal branch of DoHFuse without changing the overall architecture.

\subsection{Open World Handling}
In addition to the closed world classification task, we extend our approach to the open world setting, where the classifier may encounter traffic from websites outside the monitored set. To address this challenge, we adopt a probability-thresholding strategy based on the maximum softmax probability (MSP) \citep{hendrycks2017baseline}. Specifically, for each test sample, the model first computes the posterior probability distribution over all monitored classes. If the highest probability exceeds a given threshold $\tau$, the corresponding class is selected as the prediction; otherwise, the sample is labeled as \textit{unmonitored}. This mechanism effectively enables the model to reject unknown classes without requiring any additional training. In practice, we either fix $\tau=0.5$ or sweep across a range of thresholds to analyze the trade-off between precision and recall.

\section{Experiments}
\subsection{Experimental Environment}
\label{subsec:expen}

All experiments were conducted on a server equipped with 16 virtual CPU cores, 60 GiB of RAM, and an NVIDIA A10 GPU, running Ubuntu 24.04 LTS (64-bit). The software environment is based on Python 3.13.3, with the following key libraries and dependencies:

\begin{itemize}
  \item PyTorch: 2.8.0
  \item TensorFlow: 2.20.0rc0
  \item Scikit-learn: 1.7.1
  \item CUDA: 12.8 (Build V12.8.93)
  \item cuDNN: compatible with CUDA 12.8
\end{itemize}

To facilitate reproducibility, we have made our complete codebase publicly available\footnote{\url{https://github.com/grasstractor/DoHFuse}}.

\subsection{Limitations of Length-Based Features in DoH/3}

Among prior studies, \citep{bushart2020padding} is the most closely aligned with our setting in that it de-emphasizes length information. That said, their assumptions and traffic model differ from DoH/3, where padding and HTTP/3-over-QUIC multiplexing reshape the available signal and complicate direct transfer of length-centric pipelines. In contrast, most other approaches treat packet length as a core discriminative feature, and heavily exploit it for classification. However, as highlighted by \citep{siby2020encrypted}, once packet length information is obfuscated or padded, the classification accuracy of such models drops significantly. 

To illustrate the limitation of length-based features under DoH/3, we conducted a sanity check using a widely adopted packet length representation scheme. Based on the packet filtering strategy described in Section~\ref{sec:feature}, we extracted DoH/3-related request and response packets and constructed input sequences based on their packet lengths. Each packet is encoded by its size and signed by direction (positive for client-to-server, negative for server-to-client), a representation commonly used in existing works. We then evaluated three standard classifiers, Random Forest (RF), LSTM, and CNN, which are frequently adopted or extended in previous studies \citep{zou2021depl,dahanayaka2022inline,trevisan2023attacking,shao2023lightweight,salari2025privacy}.

The results, as summarized in Table~\ref{tab:length-degraded}, reveal a notable degradation in classification performance across all models. While Random Forest achieved an accuracy of 67.39\%, the result is far from satisfactory for a closed world setting with 449 classes. More critically, both LSTM and CNN models performed extremely poorly, yielding accuracy scores below 10\%—barely above random guessing. This suggests that deep models, which typically rely on sequential patterns in input features, failed to extract meaningful temporal signals from padded and encrypted packet length sequences under DoH/3.

These findings confirm that packet size alone is no longer a reliable signal in the DoH/3 setting, due to the protocol's encryption and obfuscation mechanisms. Consequently, relying solely on direction-signed length features leads to severe performance degradation, motivating the need for more robust temporal representations and multimodal fusion strategies.

\begin{table}[h]
\centering
\caption{Performance of common models on DoH/3 packet length sequences (direction-signed).}
\label{tab:length-degraded}
\begin{tabular}{lcccc}
\toprule
\textbf{Model} & \textbf{Accuracy} & \textbf{Precision} & \textbf{Recall} & \textbf{F1 Score} \\
\midrule
RF & 67.39\% & 68.74\% & 67.40\% & 66.07\% \\
LSTM          & 8.95\% & 2.80\% & 8.93\% & 3.73\% \\
CNN           & 7.43\% & 2.02\% & 7.39\% & 2.74\% \\
\bottomrule
\end{tabular}
\end{table}

\subsection{Closed World Comparison}

To evaluate the scalability and robustness of existing website fingerprinting techniques under HTTP/3-based DNS-over-HTTPS (DoH/3), we re-implement five representative models sourced from the recent literature~\citep{bushart2020padding,shao2023lightweight,dahanayaka2022inline,csikor2025dns,salari2025privacy}. All models preserve their original architectures and training strategies. The sole adaptation concerns the input representation: each trace is mapped to a timing-only format as a sequence of inter-request intervals $\Delta t$, allowing fair and consistent comparison under the \emph{length-obfuscated} setting that DoH/3 enforces via QUIC padding and EDNS(0) behavior. The specific adaptations applied to each baseline model are summarized as follows.

\textbf{(1)k-Nearest Neighbors.} Bushart and Rossow~\citep{bushart2020padding} proposed a KNN-based classifier operating on symbolic ``DNS sequences'' that jointly encode padded message sizes and inter-response delays, with edit distance as the similarity metric. In our adaptation, only timing information is preserved. Each trace is represented as a fixed-length vector of inter-request intervals (padding values set to $-1$), and Euclidean KNN (default $k=1$) is applied. A grid search on $k\in\{1,3,5,7,9\}$ is also performed. This yields a non-parametric, timing-only baseline that reflects temporal proximity between traffic traces.

\textbf{(2) Random Forest with time $n$-grams.} Shao et al.~\citep{shao2023lightweight} introduced a Random Forest classifier that uses bag-of-$n$-gram features constructed over packet-length tokens. We preserve their methodology but adapt it to timing: each inter-request interval $\Delta t$ is multiplied by a scale factor (e.g., 10\,ms), rounded, and mapped to discrete tokens. Unigram and bigram vocabularies are built on the training set, and token count per-sample histograms serve as input to a Random Forest with 70 trees, Gini impurity, and subsampling of features $\sqrt{d}$. This preserves the lightweight and interpretable design of the original while aligning with the DoH/3 timing-only constraint.

\textbf{(3) LSTM with masked timing sequences.} Dahanayaka et al.~\citep{dahanayaka2022inline} proposed a recurrent model using two stacked LSTM layers with 128 units and 0.2 dropout, designed for inline DoH fingerprinting. Their input format is a 1-D signed packet-length sequence. In our version, the architecture and training pipeline (Adam optimizer, class-balanced cross-entropy, early stopping validation) are retained, but the input is replaced with masked sequences $\Delta t$ (padding represented by $-1$). Masking is handled via built-in recurrent layer support, preserving variable-length behavior.

\textbf{(4) Transformer-based timing classifier.} Csikor et al.~\citep{csikor2025dns} proposed the first Transformer-based model for DNS web fingerprinting, using a ViT-style encoder operating on per-packet embeddings that include direction, packet length, timing, size ratios, and Δ timestamp. Their published code supports both closed-world and open-world classification using a confidence threshold on softmax probabilities.
Among all baselines, this is the only one with publicly available code, which we directly reuse, adjusting only the input embedding: each DoH/3 trace is represented as a $(L,1)$ tensor of inter-request intervals, padded to length $L=200$. We preserve their use of a learnable "SLA token", sinusoidal positional encodings, two Transformer Encoder layers (8 heads, 256 ff-dim), and a multi-layer perceptron (MLP) head.

\textbf{(5) Hybrid FCNN+GRU ensemble.} Salari et al.~\citep{salari2025privacy} introduced a hybrid ensemble architecture that combines a fully-connected network over flow-level statistics, a CNN over packet-length sequences, and a GRU over timing for \\ query/response pairs. Noting that packet-length sequences lose discriminative power under HTTP/3 padding (see Table~\ref{tab:length-degraded}), we remove the CNN branch while retaining the FCNN (statistical features) and GRU (timing features) components. Predictions are fused via soft voting, matching the original ensemble principle while discarding length-only structure. 

All models are run under the same preprocessing pipeline, with trace length padded to $L=200$ and inter-request intervals normalized when necessary. The hyperparameters for each baseline follow their respective original publications whenever applicable. For components not fully specified in the original works, we adopt a unified tuning strategy: model parameters are selected based on performance on a held-out validation set under the same data split and preprocessing conditions.

For clarity, we note that only the implementation by Csikor et al. is publicly available, which we directly reuse with minimal modification to the input representation. The remaining baselines are re-implemented according to their methodological descriptions. To ensure fairness, all models are trained and evaluated under identical data splits, preprocessing steps, and evaluation metrics. No model-specific optimization beyond validation-based tuning is applied.

% \begin{figure*}[ht]
%   \centering
%   % 第一行
%   \begin{subfigure}[t]{0.45\textwidth}
%     \includegraphics[width=\textwidth]{accuracy_vs_classes.png}
%     \caption{Accuracy vs. number of classes}
%   \end{subfigure}
%   \hfill
%   \begin{subfigure}[t]{0.45\textwidth}
%     \includegraphics[width=\textwidth]{precision_vs_classes.png}
%     \caption{Precision vs. number of classes}
%   \end{subfigure}

%   % 第二行
%   \begin{subfigure}[t]{0.45\textwidth}
%     \includegraphics[width=\textwidth]{recall_vs_classes.png}
%     \caption{Recall vs. number of classes}
%   \end{subfigure}
%   \hfill
%   \begin{subfigure}[t]{0.45\textwidth}
%     \includegraphics[width=\textwidth]{f1-score_vs_classes.png}
%     \caption{F1-score vs. number of classes}
%   \end{subfigure}
  
%   \caption{Performance metrics of different models under varying classification granularities (100, 200, 449 classes).}
%   \label{fig:line-metrics}
% \end{figure*}

\begin{figure*}[ht]
  \centering
  \includegraphics[width=1\linewidth]{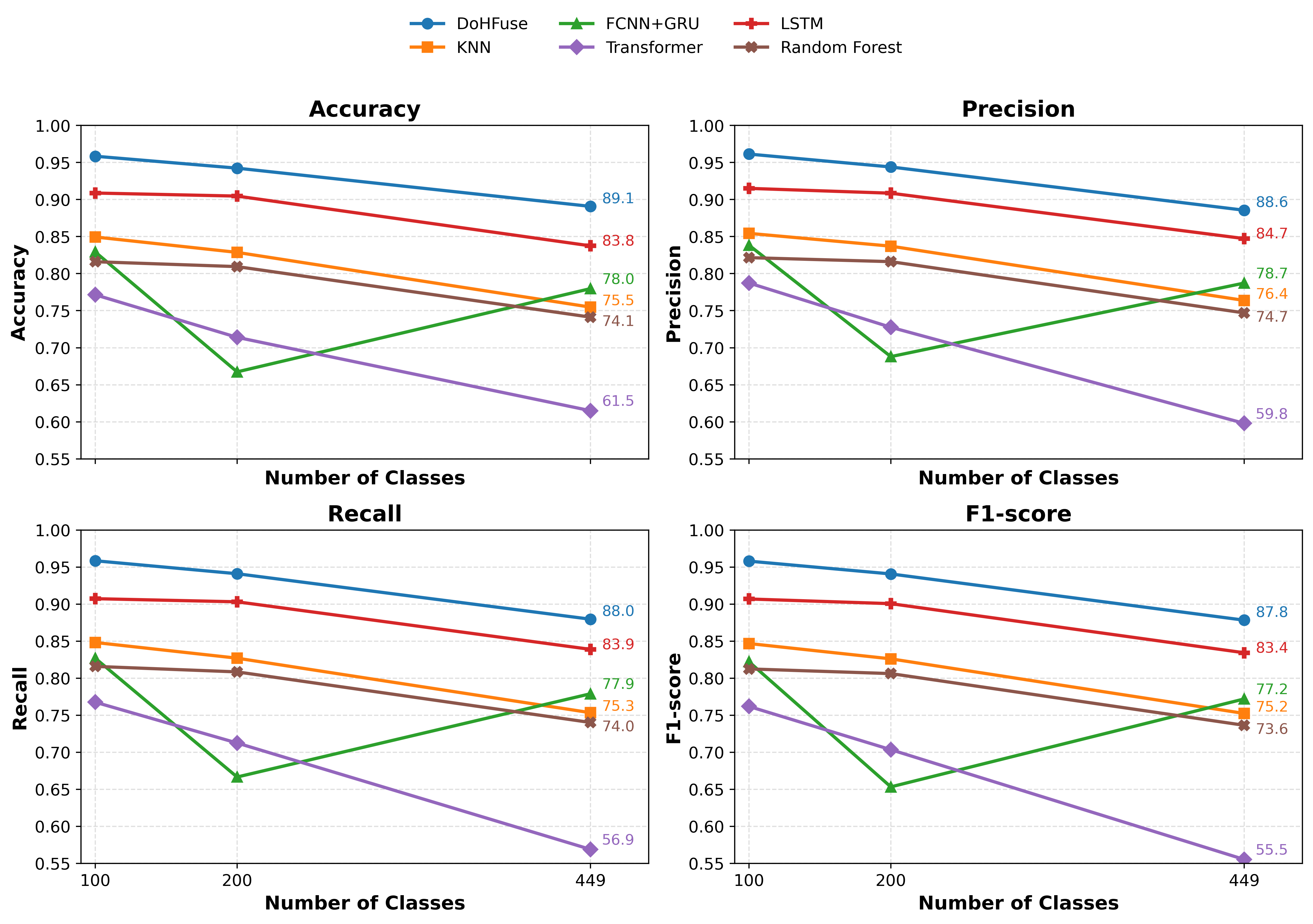}
  \caption{Performance metrics of different models under varying classification granularities (100, 200, 449 classes).}
  \label{fig:line-metrics}
\end{figure*}

As shown in Figure~\ref{fig:line-metrics}, we compare the performance of all methods across four widely used metrics: accuracy, precision, recall, and F1-score. In general, the performance of all models declines as the number of classes increases, which is expected because the classification task becomes progressively more difficult. Nevertheless, the relative ranking of the methods remains consistent across metrics.

Among the compared baselines, models that emphasize sequential temporal modeling generally achieve better performance than traditional statistical or shallow methods. In particular, LSTM is the strongest timing-only deep baseline across all settings, while KNN and Random Forest still retain reasonable discriminative ability as the number of classes grows. In contrast, the vanilla Transformer consistently underperforms, suggesting that a purely attention-based architecture is less effective at capturing the fine-grained temporal dependencies embedded in one-dimensional DoH/3 interval sequences. 

A particularly noteworthy case is FCNN+GRU. Although it remains competitive in the simpler setting, its performance drops sharply in the 200-class scenario. This suggests that the statistical features processed by the FCNN branch are not sufficiently discriminative when class overlap increases, and that assigning equal importance to statistical and temporal cues may lead the model to overemphasize noisy or weakly informative inputs.

Overall, our proposed DoHFuse consistently outperforms all baselines under every class configuration and across all evaluation metrics. The performance gap becomes especially pronounced in the full 449-class setting, where the competing methods suffer more severe degradation, while DoHFuse maintains strong classification performance. We attribute this advantage to two main factors: (1) the dual-branch design effectively integrates temporal and statistical views through a balanced fusion strategy, rather than relying on either one in isolation; and (2) the DMAG-LSTM module is better suited to modeling the burst--gap temporal patterns that become increasingly important as the classification granularity grows.

While Figure~\ref{fig:line-metrics} presents representative results, a single run does not fully reflect the statistical stability of each method. To provide a more comprehensive evaluation, we further repeat the key closed-world experiments under multiple random seeds and summarize the aggregated results in Table~\ref{tab:main_stat}.

\begin{table*}[!ht]
\centering
\caption{Statistical comparison of closed-world performance (mean ± std).}
\label{tab:main_stat}
\resizebox{\textwidth}{!}{
\begin{tabular}{l c c c}
\hline
Model & CW-100 Accuracy (mean $\pm$ std) & CW-200 Accuracy (mean $\pm$ std) & CW-449 Accuracy (mean $\pm$ std) \\
\hline
KNN & 84.95 $\pm$ 0.88 & 82.87 $\pm$ 1.67 & 75.48 $\pm$ 1.12 \\
Random Forest with time $n$-grams & 81.61 $\pm$ 0.73 & 80.95 $\pm$ 0.91 & 74.13 $\pm$ 0.95 \\
FCNN+GRU & 82.93 $\pm$ 1.12 & 66.74 $\pm$ 2.36 & 77.98 $\pm$ 1.48 \\
LSTM & 89.70 $\pm$ 0.61 & 90.77 $\pm$ 0.81 & 83.75 $\pm$ 0.92 \\
Transformer & 75.93 $\pm$ 2.18 & 71.05 $\pm$ 1.08 & 61.50 $\pm$ 5.29 \\
DoHFuse & \textbf{95.48 $\pm$ 0.32} & \textbf{94.33 $\pm$ 0.14} & \textbf{89.08 $\pm$ 0.49} \\
\hline
\end{tabular}
}
\end{table*}

As shown in Table~\ref{tab:main_stat}, DoHFuse not only achieves the best average accuracy across all three class settings, but also maintains relatively small standard deviations, indicating stable optimization and strong robustness under the unified DoH/3 protocol. Among the baseline models, LSTM is the strongest and most stable timing-only deep model. In contrast, the vanilla Transformer is not only substantially weaker on average, but also noticeably less stable across random seeds, especially in the 449-class setting, where the standard deviation exceeds 5 percentage points. This indicates that a standard Transformer is not well matched to the burst--gap temporal dynamics of DoH/3 traffic in our setting.

FCNN+GRU also exhibits comparatively large variance, particularly in the 200-class setting, which is consistent with the trend observed in Figure~\ref{fig:line-metrics}. This result further suggests that statistical features alone, or overly balanced fusion without sufficiently strong temporal modeling, may become unreliable when inter-class boundaries become more complex. Traditional machine learning baselines such as KNN and Random Forest remain competitive in simpler settings, but their performance degrades more noticeably as the task becomes finer-grained. In particular, KNN shows relatively large variance in the 200-class setting, indicating greater sensitivity to neighborhood instability under increased class overlap.

Taken together, these closed-world results directly answer our motivating question: even in the presence of EDNS(0) padding and QUIC multiplexing, DoH/3 remains vulnerable to timing-based website fingerprinting. Inter-arrival structures still preserve site-specific regularities that are sufficiently strong for high-accuracy identification. Therefore, padding and multiplexing alone are not enough to eliminate website fingerprinting leakage in practice.

\begin{figure}[!ht]
  \centering
  \includegraphics[width=0.95\linewidth]{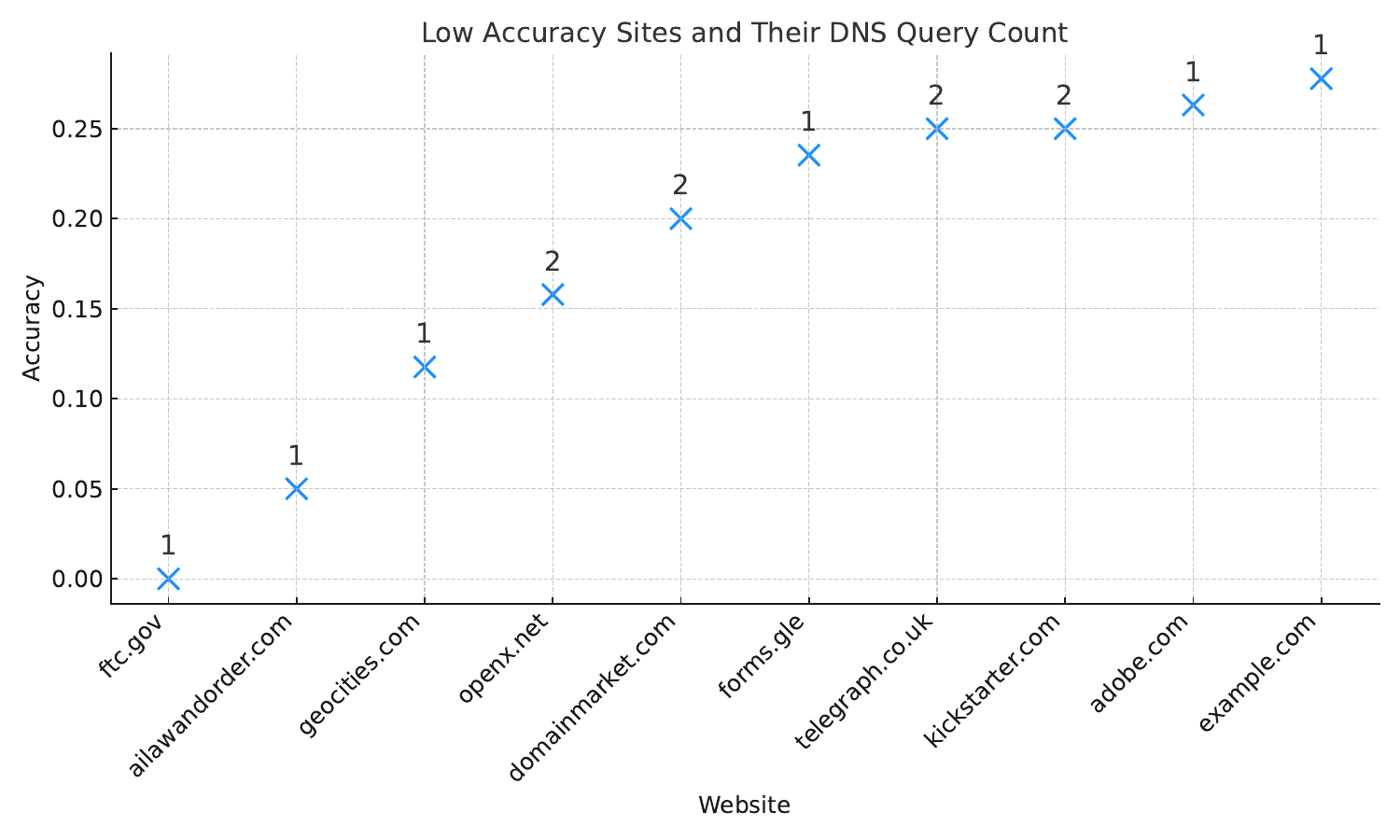}
  \caption{
    Accuracy of the least identifiable websites and their corresponding number of observable non-Google DNS queries.
    The numeric label above each point indicates the maximum number of such queries observed in a single visit.
  }
  \label{fig:fail-sites-analysis}
\end{figure}

\subsection{Failure Analysis of Hard-to-Classify Websites}
\label{sub:low-acc}

Figure~\ref{fig:fail-sites-analysis} illustrates our focused analysis on the websites that perform the lowest in our dataset - those with the poorest classification accuracy under our DoH/3 fingerprinting framework.

These domains typically generate extremely sparse DoH traffic, with only one or two observable non-Google DNS queries per visit. This lack of query activity results in highly indistinct temporal patterns, which offer an insufficient signal for the classifier to learn from.

As shown in the figure, all these sites show a classification precision below 30\%, with domains such as \texttt{ftc.gov} and \texttt{ailawandorder.com} performing near zero.

This observation highlights a fundamental limitation in current fingerprinting approaches: lightweight or static websites with minimal DNS interactions are inherently difficult to distinguish, especially over DoH/3 where padding and encryption already reduce feature visibility. The absence of burstiness or well-separated timing gaps makes accurate identification particularly challenging in these cases.

\subsection{Open World Evaluation}
\label{sec:openworld_eval}

\begin{figure}[!ht]
  \centering
  \includegraphics[width=0.9\linewidth]{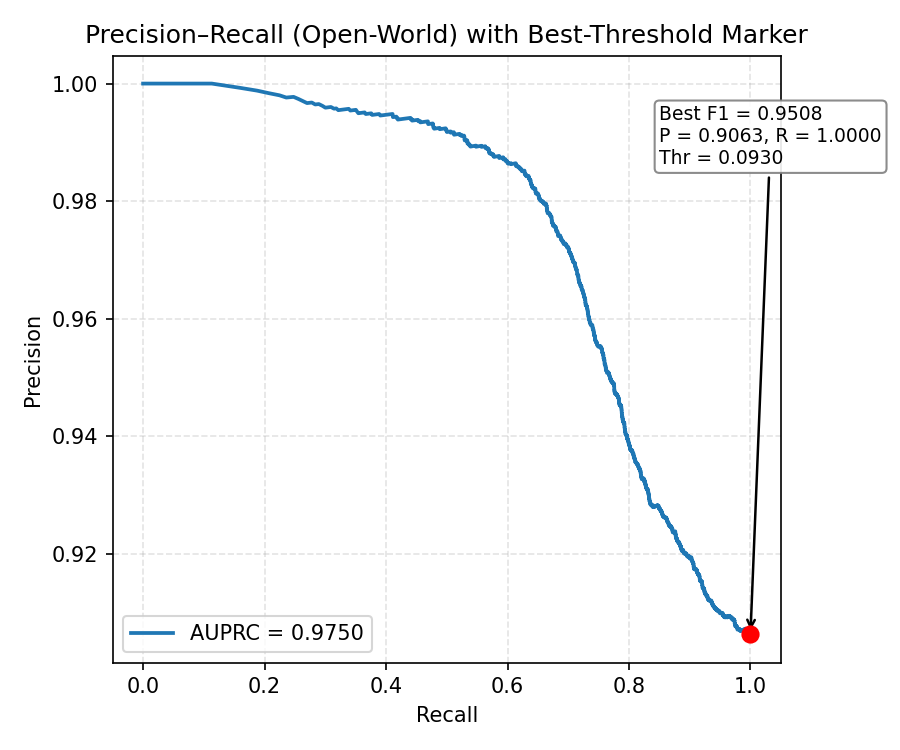}
  \caption{Precision--Recall curve of the proposed model in the open world setting. The red marker denotes the best-threshold operating point.}
  \label{fig:openworld_pr}
\end{figure}

To further assess the practicality of our proposed model, we perform open world experiments in which the classifier encounters both monitored and unmonitored websites. Following prior work, we adopt the maximum softmax probability (MSP) as the confidence measure and apply a probability threshold to differentiate monitored traffic from previously unseen traffic. Specifically, if the highest predicted probability of a sample falls below the threshold, it is labeled as unmonitored.  

As shown in Fig \ref{fig:openworld_pr}, our model achieves strong performance in the open world setting. The Precision--Recall (PR) curve yields an AUPRC of 0.975, suggesting that the model reliably detects monitored websites even in the presence of large-scale unmonitored traffic. The best F1-score reaches 0.951 with a precision of 0.906 and a recall of 1.0, showing that all monitored websites are correctly identified while maintaining a relatively low false positive rate. Even when using the default threshold of 0.5, the model still achieves an F1-score of 0.918, highlighting its robustness across different operating points. These findings confirm that the proposed approach generalizes well under realistic open world conditions.  

\subsection{Ablation Study}

\begin{table*}[ht]
  \centering
  \caption{Ablation results under the full 449-class closed world setting.}
  \label{tab:ablation-results}
  \begin{tabular}{lcccc}
    \toprule
    \textbf{Model Variant} & \textbf{Accuracy} & \textbf{Precision} & \textbf{Recall} & \textbf{F1-score} \\
    \midrule
    Full Model (Ours)              & \textbf{0.8908} & \textbf{0.8856} & \textbf{0.8796} & \textbf{0.8783} \\
    w/o Statistical Features       & 0.8771         & 0.8796          & 0.8758          & 0.8732 \\
    Replace DMAGLSTM with LSTM     & 0.8542          & 0.8597          & 0.8584          & 0.8590 \\
    \bottomrule
  \end{tabular}

\end{table*}

To further investigate the effectiveness of each architectural component in our model, we performed two ablation experiments in the full 449-class closed world setting.

\textbf{(1) Removing statistical features:}  
We disable the statistical branch and retain only the temporal input processed by the DMAGLSTM module. This experiment evaluates whether the statistical branch contributes to the classification performance.

\textbf{(2) Replacing DMAGLSTM with standard LSTM:}  
We retain the dual-branch architecture but substitute our custom DMAGLSTM module with a standard LSTM layer of identical dimensionality. This isolates the impact of our temporal module design.

The comparison results are summarized in Table~\ref{tab:ablation-results}. We report four common metrics to comprehensively assess the performance of the model.

From the results, we observe that both the statistical features and the DMAGLSTM module contribute significantly to the final performance. Removing statistical features results in a moderate performance drop, confirming their value as complementary inputs. Meanwhile, replacing DMAGLSTM with a standard LSTM yields a larger degradation across all metrics, highlighting the temporal modeling strength of DMAGLSTM. These findings validate our design choices and suggest that both branches and the custom temporal module are essential to achieve robust classification in complex DoH/3 scenarios.

This ablation analysis also provides interpretability into DoHFuse’s decision process: the temporal component extracts fine-grained inter-request dynamics, while the statistical branch contributes global behavior cues that remain robust under padding or stream coalescing. The dual-branch architecture thus synthesizes local and global timing patterns, which proves crucial to classification accuracy across diverse DoH/3 scenarios.

\section{Discussion}

\subsection{Computational Overhead and Deployment Considerations.}
We briefly discuss the computational overhead of the proposed DoHFuse framework. The model consists of a dual-branch architecture combining a temporal recurrent component (DMAG-LSTM) and a lightweight fully connected network over statistical features. 

From a complexity perspective, DMAG-LSTM retains the same order of computational complexity as a standard LSTM, with only modest additional gating operations, resulting in linear time complexity with respect to the input sequence length. The statistical branch introduces negligible overhead due to its low-dimensional input and shallow structure. 

Importantly, website fingerprinting is typically performed in an offline or post-processing setting, where strict real-time constraints are not required. Therefore, inference latency is not a primary bottleneck in most practical deployments. In our implementation, inference for a single trace is lightweight and can be efficiently parallelized across batches, making the model suitable for large-scale traffic analysis scenarios.

Overall, the proposed architecture achieves a favorable balance between modeling capacity and computational cost, and is practical for real-world deployment in monitoring or analysis systems.

\subsection{Applicability to DoQ Traffic}
While our study focuses on DoH/3 (DNS-over-HTTP/3), the proposed methodology is theoretically compatible with DNS-over-QUIC (DoQ) traffic as well, since both protocols share similar bursty and encrypted communication characteristics. However, based on our observation, native DoQ traffic remains scarce in the wild \citep{kosek2022one}. Currently, there is little support from mainstream operating systems and browser engines for DoQ \citep{kosek2022dns}; prior studies that investigated DoQ fingerprinting often relied on artificial environments, such as setting up `dnsproxy` on Ubuntu \citep{hu2024privacy,salari2025privacy}. Such configurations are rarely found among real users, which limits their generalizability.

A notable exception is AdGuard, a Windows commercial security application that can intercept system-wide DNS requests and automatically encrypt them; When configured with AdGuard DNS, it can forward queries over DNS-over-QUIC (DoQ) to AdGuard’s own encrypted resolvers \citep{Bagirov2020DoQ}. In our controlled Windows measurements (with the AdGuard forwarding system DNS over QUIC to AdGuard DNS), we consistently observed nearly uniform DoQ packet sizes for queries and responses; we attribute this effect to strict EDNS(0) padding, although we are not aware of any official statement confirming such a default (background on EDNS(0) padding and padding policies). While uniformly sized packets can reduce size-based leakage, AdGuard’s broader adoption may be constrained by its closed-source model and the presence of paid subscription plans for its products and DNS service.

\subsection{Challenges from Sparse Website Traffic}
Our empirical results (see Section~\ref{sub:low-acc}) show that certain websites consistently yield very low classification accuracy. Closer inspection reveals that these sites typically generate only one or two observable DoH queries during page loading, leaving almost no meaningful temporal patterns for the model to learn from. This highlights a practical limitation of fingerprinting sparse or static websites.

Nevertheless, such cases are relatively rare. Most real-world websites, especially popular and media-rich portals, tend to have complex page structures that involve multiple third-party DNS lookups, resulting in more chaotic and distinguishable traffic patterns \citep{urban2020beyond}. Our model demonstrates robust performance across these more representative examples.

\subsection{LLMs in Website Fingerprinting}

Although large language models (LLMs) have achieved remarkable success in many sequence modeling tasks, they are currently unsuitable for DoH/3 traffic fingerprinting. Key limitations include modality mismatch (between numeric time-series data and textual language data), interference from pre-training noise, and practical difficulties in performing frequent scenario-specific fine-tuning as typically required in fingerprinting attacks.

Nevertheless, LLMs hold considerable promise for future defense mechanisms, particularly in automating the design of countermeasures. For example, their strong generative and reasoning capabilities could be leveraged to assist in designing adaptive traffic perturbation strategies, such as dynamically injecting dummy queries or adjusting packet intervals, while respecting operational constraints and countering specific attacker models.

Therefore, while this work prioritizes lightweight time series models to address immediate challenges, we encourage future research to explore (i) developing time-series or multimodal foundation models tailored to encrypted traffic analysis and (ii) focusing on the potential of LLMs to aid in the synthesis of defense policies and adaptive traffic shaping.

\subsection{Limitations and Future Work}
Although our study demonstrates that DoHFuse remains robust under DoH/3 traffic, several limitations deserve discussion.

First, our analysis primarily focuses on the timing features extracted from Chrome, the only major browser currently supporting DoH/3 natively using its built-in QUIC stack. We further verified that Chrome exhibits consistent DoH/3 resolution behavior across Windows, macOS, and Linux, indicating that operating system differences do not alter the timing characteristics used in our study. Other browsers such as Firefox (which currently relies on DoH/2) and Safari (which requires non-standard system-level configuration and does not provide practical DoH/3 deployment) were therefore not included, limiting cross-client generalizability. Moreover, prior empirical studies have shown that encrypted DNS mechanisms do not impose significant CPU or memory overhead under normal browsing conditions\citep{botter2019,Hounsel2020}, and DoH/3, based on QUIC, is expected to introduce even less performance interference, suggesting that client resource availability is unlikely to meaningfully distort timing features except under extreme load.

Second, our closed-world setting treats each full-page visit as a distinct classification target. However, certain domains may have rich hierarchical structures where visiting a deep link (e.g., store.example.com/product/) triggers a markedly different DNS query pattern from the homepage. Although these are currently treated as “different websites,” further work is needed to explore structured labels or hierarchical site modeling to better capture such real-world behaviors.

Third, although our dataset is collected from two geographically distinct vantage points and covers a large selection of sites, we observed that some websites exhibit content-level variations across regions (e.g., CDN-driven resource changes). This suggests that location-aware adaptation might be required for real-world attacks, posing practical challenges for large-scale deployment of timing-based fingerprinting models.

Fourth, our study does not explicitly evaluate robustness under active defensive countermeasures such as randomized padding, traffic shaping, or cover traffic. While such mechanisms can potentially reduce the distinguishability of timing patterns, they often introduce non-negligible bandwidth and latency overhead. In practice, unless combined with stronger obfuscation strategies, residual timing structures may still remain observable. A systematic evaluation of these defense mechanisms would require a substantially expanded experimental framework and is left for future work.

Despite these limitations, the findings highlight the persistent risks associated with timing side channels in encrypted DNS traffic. Future efforts could explore browser-agnostic modeling, hierarchical classification schemes, defense-aware evaluation, and extensive global data collection to further enhance robustness and generalizability.

\section{Conclusion}
This work establishes that website fingerprinting (WF) remains feasible under DoH/3 even when packet sizes are obfuscated by encryption, padding, and QUIC multiplexing. Theoretically, our analysis shows that the decisive signal in DoH/3 lies in timing specifically, the burst-gap request dynamics - rather than in packet lengths. Practically, these findings mean that DoH/3 should not be assumed to deliver anonymity by default: effective defenses must explicitly target timing leakage (e.g. shaping that disrupts burst regularity) and be evaluated against timing-aware adversaries, not just length-based ones.

Methodologically, we contribute a DoH/3-specific perspective that differs from most prior WF studies centered on Tor/HTTPS or on DoH/DoT without robust padding. We release a standardized DoH/3 benchmark and a unified protocol for same dataset comparisons; and we introduce DoHFuse, a dual-branch architecture that fuses inter-arrival timing with compact statistics. Its DMAG-LSTM encoder embodies the required multi-scale gating and aligns with DoH/3’s burst-aligned request patterns. Ablations and baseline comparisons indicate that length-centric pipelines do not transfer cleanly to DoH/3, whereas timing-first fusion provides robust closed world and open world behavior.

Beyond academic interest, this work sheds light on practical scenarios where DoH/3 traffic analysis may assist security and performance management. For instance, enterprise networks or national-scale CERTs may deploy passive monitoring tools to detect anomalous DoH/3 patterns that suggest malware activity or domain fronting. Similarly, content delivery providers and DNS resolvers may leverage traffic fingerprinting to optimize caching and prefetching schemes tailored to specific website access behaviors.

\section*{Acknowledgments}
This work is supported by the Shandong Provincial Natural Science Foundation(Grant No. ZR2024QF138), the S\&T Program of Shijiazhuang (Grant No. XXXX)
, and the Scientific Research Innovation Fund of Harbin Institute of Technology [Grant No. IDGAZMZ00210335].

\ifCLASSOPTIONcaptionsoff
  \newpage
\fi



%\bibliographystyle{IEEEtran}

\bibliographystyle{IEEEtran} % Elsevier 推荐样式之一（作者-年份）
\bibliography{IEEEexample}    


%\vfill


\end{document}